\documentclass[12pt,preprint]{aastex}

\def\hide#1{}

\begin{document}
\title{Kilohertz QPO Frequency Anti-Correlated with mHz QPO Flux in 4U 1608-52}
\author{Wenfei Yu\altaffilmark{1,2} and Michiel van der Klis\altaffilmark{1}}
\altaffiltext{1}{Astronomical Institute, ``Anton Pannekoek'', 
	University of Amsterdam, and Center for High Energy
	Astrophysics, Kruislaan 403, 1098 SJ Amsterdam, The
	Netherlands. E-mail: yuwf@astro.uva.nl; michiel@astro.uva.nl}

\altaffiltext{2}{Laboratory for Cosmic Ray and High Energy Astrophysics, 
	Institute of High Energy Physics, Beijing, 100039, China (on leave) }

\begin{abstract}
We analysed {\it Rossi X-ray Timing Explorer} (RXTE) data of the
low-mass X-ray binary and atoll source 4U 1608-52 obtained on March 3,
1996 in which the source simultaneously showed a strong single
kilohertz quasi-periodic oscillation (QPO) around 840 Hz, and a 7.5
mHz QPO detected at energies below 5 keV.  We find that the frequency
of the kHz QPO is approximately anti-correlated with the 2--5 keV
X-ray count rate associated with the mHz QPO. The average
kHz QPO frequency varies by about 0.6 Hz (0.07\%) during a mHz QPO
cycle over which the average 2--5 keV count rate varies by about 60
c/s (4\%). This is opposite to the frequency-count rate correlation
observed in the same data on longer time scales and hence constitutes
the first example of a sign reversal in the frequency-flux correlation
related to the origin of the flux. Such a sign reversal is predicted
by the radiative disk truncation model for the case where the flux
variations originate on the neutron star but are not due to disk
accretion rate fluctuations. The results support the nuclear burning
interpretation of the mHz QPO, and the interpretation of the kHz QPO
frequency as an indicator of the orbital frequency at the inner edge
of the accretion disk. The varying radiative stresses on the inner
disk exerted by the flux due to the quasi-periodic nuclear burning
lead to changes in the inner disk radius and hence to the observed
anti-correlation between kHz QPO frequency and X-ray count rate.
There is a time lag of about ten seconds of the X-ray count rate
relative to the kHz QPO frequency in terms of the anti-correlation we
found between these two quantities, which could be caused by the
propagation of the nuclear burning front on the neutron star away from
the equatorial region. 
\end{abstract}


\section{Introduction}
Kilohertz quasi-periodic oscillations (kHz QPOs) in X-ray flux have
been observed from about 20 neutron star low mass X-ray binaries
(LMXBs) (see van der Klis 2000 for a recent review). Their frequencies
correspond to the dynamical time scale at radii of less than a few
tens of kilometers, and are usually regarded as associated with the
Keplerian orbital frequency of material at the inner disk
edge. Observations indicate that when on a time scale of hours to days
the accretion rate through the disk increases, the QPO frequency
increases as well, indicating that the inner disk edge moves
in. Evidence for this has been observed in the form of a positive
correlation on these time scales of kHz QPO frequency with X-ray count
rate for low luminosity LMXBs, i.e. 'atoll' sources (see Mendez et
al. 1999), and in the form of a similar correlation with curve length
S$_z$ along the track traced out in an X-ray color-color diagram for
high luminosity LMXBs, i.e. 'Z' sources, which in some cases
corresponds to an anti-correlation between kHz QPO frequency and X-ray
count rate (Wijnands et al. 1997; Homan et al. 2001; Yu, van der Klis
\& Jonker 2001).

The inner edge of the disk could be set by magnetosphere-disk
interaction (Strohmayer et al. 1996, Zhang, Yu \& Zhang 1998, Cui 2000,
Campana 2000), and for sufficiently compact neutron stars with sufficiently 
weak magnetic field is certainly
expected to be limited by the general relativistic marginally stable
orbital radius. However, in the sonic point model for kHz QPOs (Miller,
Lamb and Psaltis 1998, also Miller and Lamb 1995) radiative stresses
provide the dominant disk truncation mechanism. In this description,
removal of angular momentum from the flow by radiation drag, and the
radial radiation force combine to set the inner disk radius and
thereby the QPO frequency. For constant radiation, the disk moves {\it
in} when the disk mass flow increases; for constant mass flow, the
disk moves {\it out} when the radiation impinging upon the inner disk
edge becomes more intense. Usually, disk mass flow and radiation
change in concert; calculations show that under certain assumptions
the inner disk radius will then move {\it in} and hence QPO frequency
will go {\it up} when disk accretion and the resulting X-ray flux both
increase, and {\it vice versa} (Miller et al. 1998). If disk accretion
is not the only mechanism producing radiation more complicated
relations can occur (van der Klis 2001).

A test of the role of radiative stress for disk truncation can be
obtained by studying the changes in kHz QPO frequency resulting from
changes in X-ray flux that are {\it not} due to disk mass flow
changes. Two previous attempts at performing this test led to results
that were somewhat ambiguous. (i) An anti-correlation was
recently found between the kHz QPO frequency and the NBO (``normal
branch oscillation'') count rate in Sco X-1 ($\sim$22 Hz shift within
the 1/6-s NBO cycle, Yu, van der Klis \& Jonker 2000). However, Sco
X-1 was in a state where such an anti-correlation is also seen for
long-term flux changes that are presumably associated with disk flow
changes. (ii) A 37-Hz decrease of the kHz QPO frequency was observed
after an X-ray burst in Aquila X-1 (Zhang, Jahoda \& Kelley et
al. 1998; Yu, Li \& Zhang et al. 1999). However, other similar bursts did
not show this effect.

In this Letter, we show that in the atoll source 4U 1608--52 the kHz
QPO frequency is clearly anti-correlated with the X-ray count rate
variations associated with the 7.5 mHz QPO found by Revnivtsev,
Churazov \& Gilfanov et al. (2001). This QPO occurs in a narrow X-ray
luminosity range and they suggested, among other possibilities, that 
these were due to a special mode of nuclear burning on the neutron star
surface. This is the first clear-cut case of a sign reversal in the
kHz QPO frequency vs. count rate correlation related to the basic
process producing the radiation (disk accretion vs. nuclear
burning). It provides strong evidence in favor of a role for radiative
stresses in the disk truncation mechanism, as well as for the nuclear
burning scenario for the mHz QPO.

\section{Observations}
We used data of 4U 1608--52 obtained with the Proportional Counter
Array on-board the {\it Rossi X-ray Timing Explorer} on March 3,
1996. Previous analysis of this data set led to the discovery of a
narrow kilohertz QPO in this source varying in frequency in the range
825--880 Hz with an rms amplitude around 6--8\% (Berger et al. 1996).
Further work showed that this QPO is in fact the lower-frequency one
of a pair of kHz QPOs, and that on time scales of tens of minutes to
hours the frequencies of both these QPOs usually have a strong positive
correlation with X-ray count rate (M\'endez et al. 1998, 1999).
In this paper, we refer to this lower kHz QPO exclusively.
Recently, Revnivtsev et al. (2001) discovered a 7.5 mHz QPO in this
observation with a peak width of 2.6 Hz and an rms fractional
amplitude around 1.85\%. This mHz QPO is mostly evident in the energy
range below 5 keV when the total source luminosity is between 0.07 and
0.1 times the Eddington luminosity. The total X-ray count rate during
the observation was $\sim$3200 and $\sim$1600~c/s in the 2--60 and
2--5 keV bands, respectively; we use the former band for the kHz QPO
measurements and the latter for monitoring the mHz QPO. The data
consist of three separate contiguous segments covering 3400 s, 3500 s,
and 1800 s.  For the current analysis we use the {\it Event mode} data
{\it $E\_125us\_64M\_0\_1s$} providing 1/8192 s time resolution and 64
spectral channels covering the entire 2--60 keV band.

\section{Analysis and Results}
First, we calculated a series of Fourier power spectra using the 2--60
keV data re-binned to 1/4096~s. We used transform lengths of 4~s, which
resulted in 0.25~Hz resolution power spectra with a Nyquist frequency
of 2048~Hz. These were averaged in groups of three, to obtain one
average spectrum per 12~s. The kHz QPO frequency in each average
spectrum was taken to be that of the central bin of that 4.25-Hz wide
frequency interval which contained the highest integrated Fourier
power. The kHz QPO frequency can not be determined in this way for the
second half of the first segment, where the kHz QPO is weak (see
figure 2a in Berger et al. 1996).  For this reason we only used the
first 1000~s of this segment. As an example, the kHz QPO evolution and
the corresponding 2--5 keV light curve of the second segment, both at
12~s time resolution, are shown in Fig.~1.  The frequency of the mHz
QPO (7.5$\pm$2.6~mHz) corresponds to time scales between 100 and
200~s. On time scales longer than that, we find in accordance with
previous work (\S2), that there is a strong positive correlation
between kHz QPO frequency and X-ray count rate. Before further
analysis, in order to remove the effect of the variations on these
longer time scales, we detrended both time series by subtracting a
300-s running average.

The auto-correlation of the 2--5 keV light curves has a full width of
20--30 s, which represents the average duration of the individual mHz
QPO ``pulses''. With this in mind, we identified the peaks of the
individual mHz QPO pulses in the detrended 12-s resolution 2--5 keV
light curve by finding the highest bins in a 108-s window which was
stepped through the data with 12-s steps. We found 41 different maxima
in this way; two of these were closer than 108~s to the nearest peak
and were eliminated from further analysis, leaving a total of 39
identified well-separated mHz QPO pulses.

We aligned these pulses in time using their peaks as fiducial points. 
Then, we determined the average evolution
of count rate and kHz QPO frequency in the 108-s window around these 
peaks by averaging the corresponding segments of the detrended count rate and kHz QPO
frequency time series. The result is shown in Fig.~2. The average kHz
QPO frequency varies by about 0.6 Hz during the mHz QPO cycle, in
rough anti-correlation to the $\sim$55~c/s average 2--5~keV count rate
variation. On closer inspection, it turns out that the maximum average
X-ray count rate and the minimum average kHz QPO frequency are not
exactly aligned. Based on the overall appearance of the average kHz
QPO frequency variation, its minimum seems to occur about ten seconds 
before the maximum in average count rate.  In Fig.~3 we
plotted the average kHz QPO frequency vs. the average X-ray count rate
around the mHz QPO cycle. While the count rate is rising (filled
circles) the frequency decreases along a systematically lower path
than the one along which it increases again when the count rate is
falling (open squares). We did not detect a significant change in kHz
QPO amplitude around the mHz QPO cycle, with a rough 1 sigma upper
limit on the change in fractional rms amplitude of 0.4\% (the total
kHz QPO rms is $\sim$7.5\%).

\section{Discussion and Conclusion}
We have studied the evolution of the lower kHz QPO as a function of
the phase of the 7.5 mHz QPO displayed by 4U 1608--52 when its
luminosity is about 0.1 $L_{Edd}$. The average kHz QPO frequency
depends on the mHz QPO phase. It varies by about 0.6 Hz, or 0.07\% of
the average frequency, around the mHz QPO cycle, in rough
anti-correlation to the average 2--5 keV count rate, which varies by
about 60 c/s, or 4\% of the average 2--5 keV count rate (or
approximately 2\% of the entire PCA 2--60 keV count rate). We have
also found that in this anti-correlated variation the kHz QPO
frequency leads the count rate by about ten seconds. On longer
time scales (hours), the kHz QPO frequency in these same observations
was already known to be positively correlated with the X-ray count
rate and X-ray luminosity (Mendez et al. 1999), and our analysis
confirms this.  The fact that we find an {\it anti-correlation} on the
7.5 mHz QPO time scale, while simultaneous count rate changes not
related to the mHz QPO produce the usual {\it correlated} variations
in kHz QPO frequency demonstrates conclusively that another mechanism
than that responsible for the usual count rate variations underlies
the mHz QPO.

The mHz QPO (Revnivtsev et al. 2001) occurs only within a narrow range
of source luminosity and disappears after X-ray bursts. Its spectrum
is rather soft and consistent with a blackbody. For these reasons,
Revnivtsev et al. (2001) suggested that the mHz QPO could be due to a
special mode of nuclear burning on the neutron star surface, although
an origin due to processes in the accretion disk could not be ruled
out. We suggest that the simplest explanation for the anti-correlation
between kHz QPO frequency and mHz QPO count rate we report here is,
that the mHz QPO is generated inside the inner disk edge by the
mechanism proposed by Revnivtsev et al. (2001), and that the kHz QPO
frequency is related to the Keplerian frequency at the inner edge of
the disk, whose radius in turn is set by the radiative disk truncation
mechanism proposed by Miller et al. (1998; see \S1).  If instead the
mHz QPO were produced by a modulation of the disk accretion rate, a
positive correlation between kHz QPO frequency and mHz QPO X-ray flux
would be predicted, contrary to what we observe. Indeed, such a
positive correlation is seen for the longer-timescale flux variations.

If the mHz QPO indeed originates near the neutron star surface,
i.e. inside the inner disk radius, an anti-correlation between the
orbital frequency of disk material and the mHz QPO flux is a
prediction of the radiative disk truncation mechanism, as the inner
disk edge will move out when the X-ray flux impinging upon it
increases, due to both an increased radiation drag and an increased
radial radiation force (Miller et al. 1998).  Following their equation
(39) for the effect of the radial radiation force, taking the
luminosity of the neutron star as $\sim$0.1 $L_{E}$, a 2\% change in
the neutron star luminosity will introduce a change in orbital
frequency of about 0.1\%, i.e. 0.8 Hz for an orbital frequency of 830
Hz, or 1.1 Hz for an orbital frequency of 1060 Hz for the case when
the twin peak frequency separation is taken at a constant 230 Hz (see
M\'endez et al. 1998).  Radiation-drag effects will enhance the change
in the kHz QPO frequency; also the change in neutron star luminosity
will be larger than the 2\% we adopted if not all the source flux
originates from the neutron star surface (in most models, the hard
flux originates elsewhere).  So, quantitatively the theoretical
expectation is somewhat larger than the 0.6 Hz we observe. As
suggested below, a more detailed analysis of the emission pattern
caused by the nuclear burning may explain this.

The estimate of the frequency amplitude we made above was based on the
assumption that the relative change in mHz QPO flux we observe on
Earth is the same as that in the flux towards the inner disk
edge. This can not be true if the mHz QPO is caused by local
thermonuclear fires propagating on the neutron star. If the fires
originate mostly in the equatorial region, where most of the accreted
matter piles up, and then expand and propagate away, then the flux
amplitude experienced by the disk would be less than that observed,
and the radiative stresses exerted on the inner disk edge would reach
their maximum before the observed X-ray flux does. This could explain
both the lower than expected frequency amplitude and the time lag
between the X-ray count rate and the kHz QPO frequency which we
observed.

In this paper we have demonstrated that study of the correlation
between low-frequency variability and high frequency QPOs can provide
strong constraints on the origin of both variability modes. Further
similar studies of the relations between low-frequency variability
and, e.g., kHz QPOs in neutron star X-ray binaries and high frequency
QPOs in black hole X-ray binaries, could potentially provide
considerable additional insight into the nature of both QPOs and
broad-band variability in these objects. 

\acknowledgments
We would like to thank the anonymous referee for comments which 
improve the clarity of the presentation. This work was supported 
in part by the Netherlands Organization for
Scientific Research (NWO) under grant 614.051.002. 
WY would also like to acknowledge partial support from NSFC on
computing.  This work has made use of data obtained through the High
Energy Astrophysics Science Archive Research Center Online Service,
provided by the NASA/Goddard Space Flight Center.

\newpage

\begin{figure}
\plotone{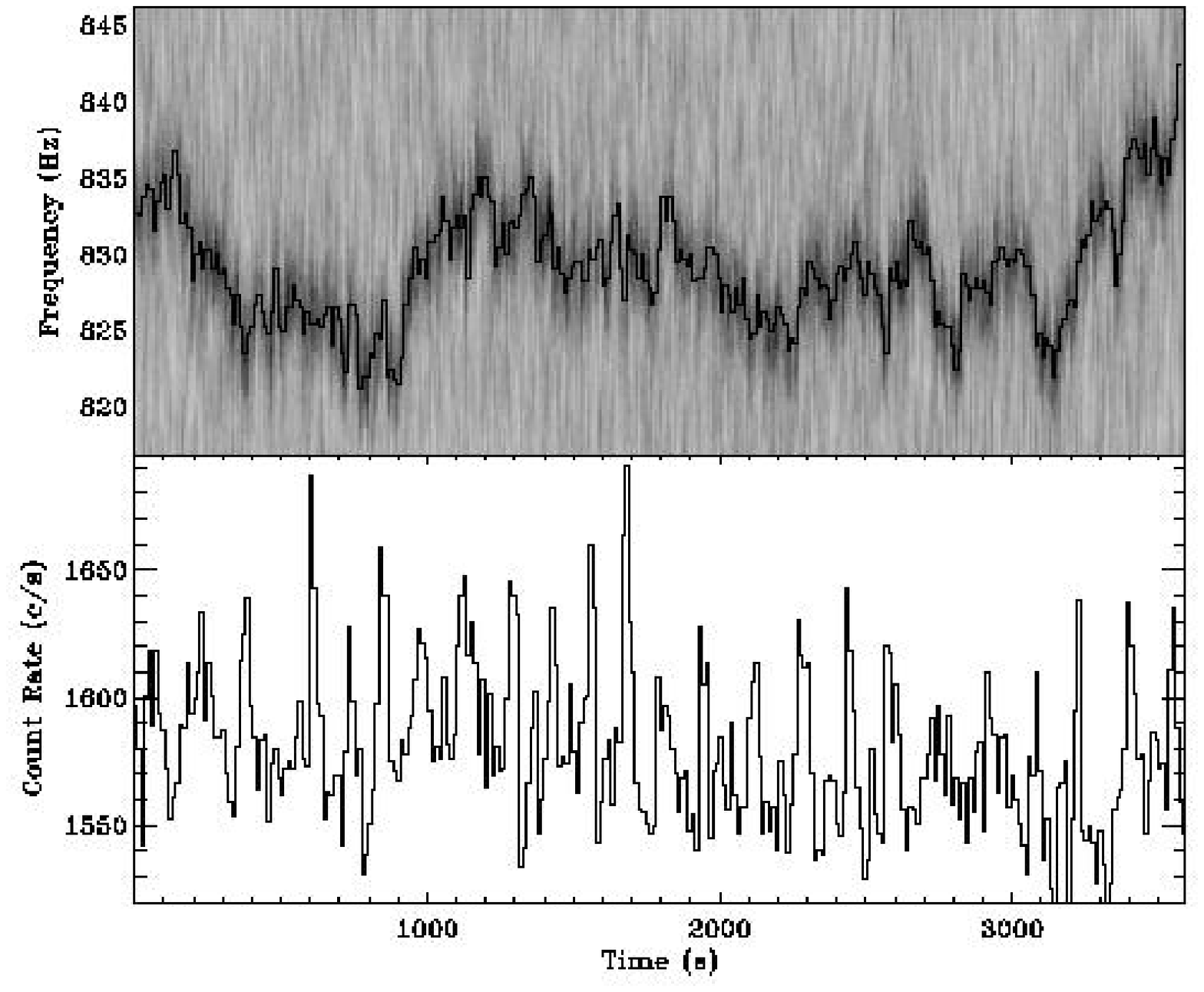}
\caption{Sample kHz QPO frequency evolution and the corresponding 
2--5 keV light curve for data segment 2. In the {\it top panel} the
grey scale indicates power; the stepped line is the kHz QPO frequency
derived as described in \S3. The time resolution in both panels is 12
s. Typical errors are about 1 Hz on the frequency estimates and about
12 c/s on the count rate estimates. }
\end{figure}

\begin{figure}
\plotone{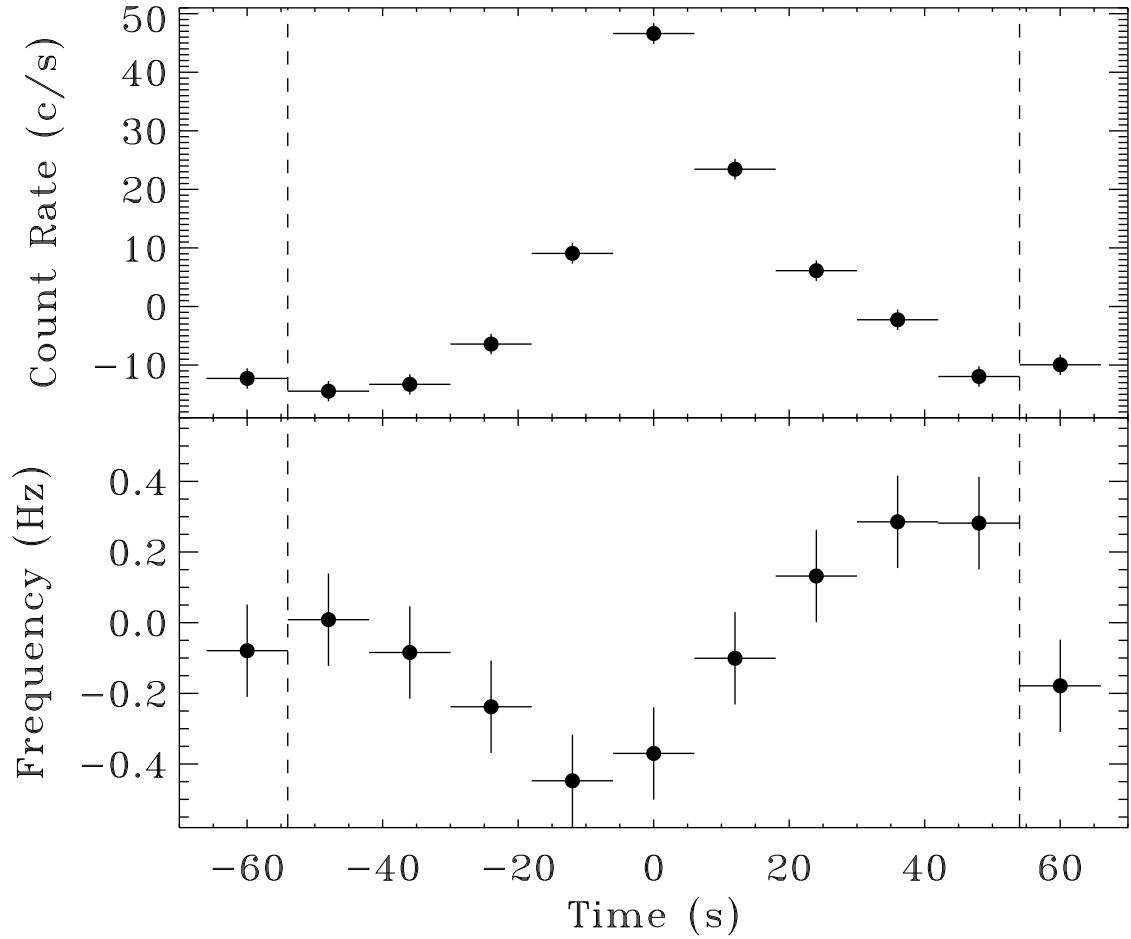}
\caption{The average mHz QPO pulse profile and 
the corresponding kHz QPO frequency evolution from the detrended 
count rate and frequency time series. The dashed lines on both sides 
mark the 108-s window within which there is no overlap with 
adjacent pulses.}
\end{figure}

\begin{figure}
\plotone{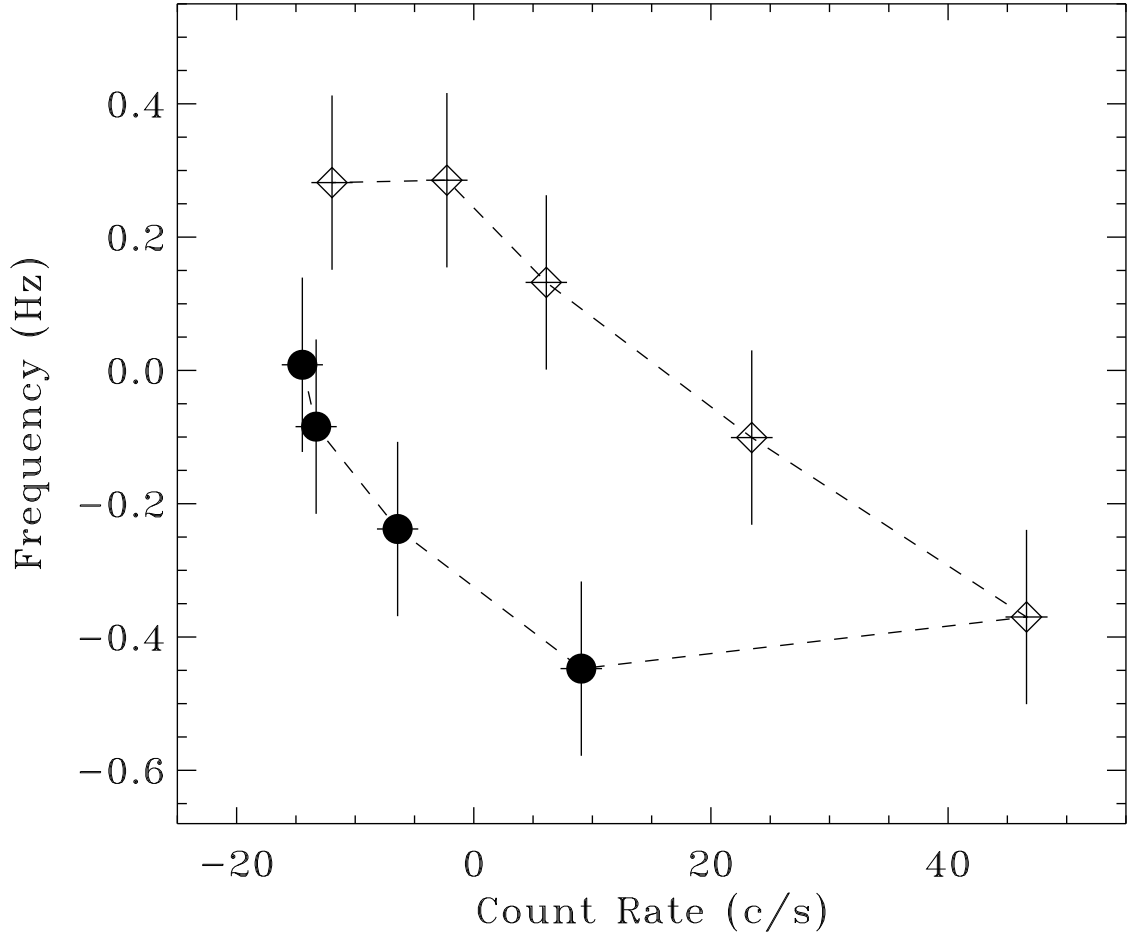}
\caption{The correlation between kHz QPO frequency and average 
count rate before and after the pulse peak of the low frequency
QPO. Data corresponding to the rise and the decay of the pulse are
marked as filled circles and diamonds, respectively. }
\end{figure}

\end{document}